\documentclass[authoryear,preprint,review,12pt]{elsarticle}

\usepackage{amssymb}
\usepackage{amsmath}
\usepackage{graphicx}
\usepackage{hyperref}
\usepackage{enumitem}
\usepackage{subcaption}
\usepackage{natbib}

\begin{document}

\begin{frontmatter}
  \date{}


\title{Optimizing AUV speed dynamics with a data-driven Koopman operator approach} 

%

\author{Zhiliang Liu\corref{cor1}\fnref{fn1}}
\cortext[cor1]{Corresponding author}
\ead{liuzhiliang@qdu.edu.cn}

\author{Xin Zhao\fnref{fn2}}
\author{Peng Cai\fnref{fn2}}

\author{Bing Cong\fnref{fn1}}
\ead{15886201166@163.com}
\affiliation[fn1]{%
  organization={School of Automation, Qingdao University},
  city={Qingdao},
  postcode={266071},
  country={China}
}
\affiliation[fn2]{%
  organization={PLA Naval Submarine Academy}, 
  city={Qingdao}, 
  postcode={266199},
  country={China}
}
\begin{abstract}
  Autonomous Underwater Vehicles (AUVs) play an essential role in modern ocean exploration, and their speed control systems are fundamental
  to their efficient operation. Like many other robotic systems, AUVs exhibit multivariable nonlinear dynamics and face various constraints, 
  including state limitations, input constraints, and constraints on the increment input, making controller design challenging 
  and requiring significant effort and time. This paper addresses these challenges by employing a data-driven Koopman operator theory combined 
  with Model Predictive Control (MPC), which takes into account the aforementioned constraints. The proposed approach not only ensures 
  the performance of the AUV under state and input limitations but also considers the variation in incremental input to prevent
  rapid and potentially damaging changes to the vehicle's operation. Additionally, we develop a platform based on ROS2 and Gazebo 
  to validate the effectiveness of the proposed algorithms, providing new control strategies for underwater vehicles against the complex and dynamic nature of underwater environments.
\end{abstract}


\end{frontmatter}

\section{Introduction}
AUVs play an essential role in modern ocean exploration, which are capable of performing a variety of tasks, including but not limited to
 underwater surveying, detection, mapping of seafloor topography, collection of geological and hydrological 
 data, as well as potential military applications. The development of AUV technology has significantly progressed over time, with modern AUVs featuring six degrees
of freedom and capable of traveling at speeds exceeding 20 meters per second \citep{sahoo2019advancements}. These advancements have made AUVs more
compact and less expensive, yet achieving full autonomy for these systems remains a significant challenge\citep{zhang2023autonomous,gafurov2015autonomous}.

 AUVs have the capability to control their position and attitude in six degrees of freedom (DOFs). The 6-DOFs equation of motion 
 for these vehicles often be partitioned into three relatively independent subsystems: speed control, longitudinal dynamics 
 (encompassing parameters such as surge, heave, and pitch), and lateral dynamics (incorporating variables like sway, roll, and yaw) 
 \citep{healey1992slow}. This decomposition is particularly advantageous for slender symmetrical bodies, such as aircraft, missiles, 
 and submarines \citep{gertler1967standard}.

This paper primarily focuses on the speed control system, which regulates the speed of underwater vehicles by controlling the rotational speed of the propeller.
The relationship between the propeller’s rotational speed and the vehicle’s speed is highly nonlinear, posing significant challenges in modeling 
and control. 
To resolve the difficulties, many thruster models have been proposed.
 \citet{yoerger1990influence} provide valuable insights into the behavior and control of torque-controlled thrusters, offering potential strategies for 
improving their performance and reliability. 
\citet{healey1995toward} formulate a two-state model that incorporates tinfoil propeller hydrodynamics, utilizing sinusoidal functions for lift and drag.
\citet{kim2006accurate} propose an accurate and practical thrust modeling for underwater vehicles that considers the effects of 
ambient flow velocity and angle. 
\citet{bachmayer2003adaptive} utilize a least-squares regression method to interpret the experimental data, aiming to optimally adjust the model 
parameters for an accurate fit. \citet{di2017effect} study the influence of ocean currents on the closed-loop performance of thrusters, considering both the dynamics of the thrusters and the 
impact of ocean currents.

Existing methods often simplify this nonlinear system into a linear one, which may not fully capture the system’s complexities. 
Moreover, obtaining an accurate model of the system can be challenging in practical engineering applications, adding complexity
 to the implementation of nonlinear control methods\citep{gong2021lyapunov,shen2017trajectory,yan2023robust}. Additionally, 
 some approaches without a specific model 
 have not completely considering critical system constraints, such as the propeller’s speed limit and the increment input limit\citep{lin2023depth,yan2019robust}.
 
The Koopman operator theory offers a compelling approach to addressing the non-linearity inherent in the speed control of AUVs. 
Unlike traditional linearization techniques, the Koopman method provides a linear representation of nonlinear dynamical systems by lifting 
the original state space into a higher-dimensional space where linear analysis can be applied\citep{mezic2005spectral,kutz2016dynamic}. 
Dynamic Mode Decomposition (DMD) is a data-driven method for analyzing dynamical systems.
The original DMD method primarily addresses linear systems\citep{rowley2009spectral,schmid2010dynamic}. To analyze nonlinear systems, extensions such as Extended DMD (EDMD) 
and Higher-order DMD have been developed. \citet{williams2015data} provide a data-driven approximation of the Koopman operator, 
extending DMD's capabilities to capture nonlinear dynamics.
This results in a more accurate and comprehensive 
model of the system dynamics, which is particularly beneficial for capturing the complex behaviors of AUVs.

Model Predictive Control (MPC) complements the Koopman method by providing a robust framework for handling constraints and optimizing
control inputs over a finite horizon\citep{korda_linear_2018,zhang2022robust,bruder2020data}. MPC's ability to anticipate future events and adjust control actions accordingly
 makes it well-suited
for managing the constraints of AUV speed control systems. By integrating the Koopman operator with MPC, 
we can leverage the strengths of both methods: the Koopman method's enhanced modeling accuracy and MPC's predictive and constraint-handling
capabilities.

This study's key contributions are as follows:
\begin{enumerate}[label=\protect\raisebox{0.25ex}{\textbullet},leftmargin=*]
  \item This paper introduces a speed control system developed by integrating Koopman operator and MPC methods, designed to enhance the accuracy 
  and robustness of AUV speed control;
  \item The proposed control system takes into account the practical constraints on system input, states, and increment input, which allows for precise tracking of the AUV's trajectory;
  \item A simulation platform that integrates Gazebo and ROS2 has been designed to verify the effectiveness of the proposed solution, 
  enabling comprehensive testing in a controlled environment and reducing the risk of mission failure prior to real-world application.
\end{enumerate}

The structure of this paper is outlined as follows. Section 2 presents the velocity model for AUV. 
Section 3 introduces the theory of the Koopman operator and details the process for achieving a high-dimensional linear space. 
Section 4 discusses the integration of Koopman operator theory with MPC for its application in AUV. 
Section 5 presents two simulation experiments, one conducted in MATLAB and the other in the physical simulation software Gazebo. 
The final section is dedicated to a discussion of the conclusions drawn from the research.

{\bf Notation:} $\mathbb{C}$, $\mathbb{R}$ and $\mathbb{Z}$ denote complex number, real number and integers, respectively. $A^{T}$, $\| A\|_F$ and $\| A\|_2$
denote transpose, Frobenius norm and Euclidean norm, respectively. 
Given matrices $A\in \mathbb{C}^{m\times n}$ and $B \in\mathbb{C}^{m\times d}$, we denote by $[A, B] \in \mathbb{C}^{m\times (n+d)}$ 
the matrix created by concatenating $A$ and $B$.
$\mathcal{T} u$ denotes next input, $\mathcal{H}$ denotes Hilbert space. $\min$ and $\max$ denote the minimum and maximum
value, respectively.
\section{Problem formulation and preliminaries}
In general, the force and moment vector of the thrust will be a complex function\citep{fossen2011handbook}. This relationship can be expressed as:
\begin{equation}
    \tau=f(v,u),
\end{equation}
where  $v\in\mathbb{R}^6$ is the underwater vehicle's velocity vector, $u\in\mathbb{R}^p(p\geq 6)$ is the control variable, 
and $f$ is a nonlinear vector function. The thrust $T$ of a single-screw propeller can be obtained by the following formula:
\begin{equation}\label{formulaT}
  T=\rho D^4K_T(J_0)|s|s  ,
\end{equation}
where $s$  represents the propeller RPM(revolutions per minute), $D$ represents the propeller diameter, $\rho$ represents the water density, 
and $V_a$ represents the propeller advance velocity (the velocity of the water flow entering the propeller). $J_0=V_a/(sD)$ is an advanced 
number and $K_T$ is the thrust coefficient. In general, the thrust in the forward and backward directions is not symmetrical. However,
 many underwater vehicle thruster systems are designed to provide symmetrical thrust. This parameter $K_T$ typically exhibits linear behavior 
 in $J_0$, so the following approximation holds: 
\begin{equation}\label{KT}
  K_T=\alpha_1+\alpha_2\frac{V_a}{sD}.
\end{equation}
In this case,  $\alpha_1$  and $\alpha_2$ are two constants.  And then, the thrust can be described as
\begin{equation}\label{T1}
  T(s,V_a)=T_{|s|s}|s|s+T_{|s|V_a}|s|V_a,
\end{equation}
where $T_{|s|s}>0$ and $T_{|s|V_a}<0$  are design parameters that depend on factors such as propeller diameter, 
duct shape, and water density. The above coefficients also depend on $s$ and $V_a$, since the above two equations are only first-order approximations 
to more general expressions. However, experiments have shown that this dependence is negligible under most practical operating conditions. 
Forward speed relation to vehicle speed can be expressed as
\begin{equation}\label{Va}
  V_a=(1-\omega)V,
\end{equation}
where  $\omega$ is the wake fraction (typically $0.1-0.4$). The propeller force generated by a single propeller can be described by a nonlinear 
function according to the results of (\ref{T1})
\begin{equation}\label{tau1}
  \tau=b_1|s|s-b_2|s|v,
\end{equation}
where $b_1=T_{|s|s}>0$,$b_2=-T_{|s|V_a}(1-\omega)$.

{\bf Remark 1:} In the multivariate case, Equation (\ref{T1}) can be rewritten as: $\tau=B_1u-B_2(u)v$ , 
  where $B_1$ and $B_2$ are matrices with appropriate dimensions, and $v$ is a newly defined control variable, expressed as 
  $u_i=s_i|s_i|$. As referenced in \cite{fossen2011handbook}, bilinear models are often approximated by simpler affine models
  (i.e., $\tau=Bu$) due to the limited theoretical framework for non-affine control systems. To address this, this study proposes 
   applying Koopman theory and MPC method for direct control of non-affine systems.
   \subsection{Velocity control model}
      According to \cite{healey1992slow},  the 6-DOFs equations of motion for an underwater vehicle 
     can be divided into three non-interacting subsystems for speed control, steering, and diving. This paper focus on the speed control.
     Assuming negligible coupling between sway, heave, roll, pitch, and yaw motions, the speed equation can be expressed as:
     \begin{equation}\label{forwardspeed}
     (m - X_{\dot{v}})\dot{v}=X_{|v|v}|v|v+(1-t)T+x_{e},
     \end{equation}
     where $m$ is the vehicle mass, $u$ is the vehicle velocity, $X_{|v|v}$ is hydrodynamic parameter, $T$ is thrust,$x_{e}$ is external disturbance, 
     $X_{\dot{v}}$ is the added inertia, $t$ is the thruster deduction number. 
     Combined equation (\ref{forwardspeed}) and (\ref{T1})(ignoring the external disturbance $x_e$), one has
   \begin{equation}\label{forwardspeed1}
     (m - X_{\dot{v}})\dot{v}=X_{|v|v}|v|v+(1-t)(T_{|s|s}|s|s+T_{|s|V_a}|s|V_a)
   \end{equation}
   
     The model under study is nonlinear, differing from the dynamics described in \citet{fossen2011handbook}. Consequently, traditional linear control methods 
     are not applicable. 
     In this study, we employ Koopman theory to derive a linear model in a higher-dimensional space. Subsequently, we use the MPC
     approach to design the control strategy.
     Equation (\ref{forwardspeed1}) represents a continuous dynamical system. In practical applications, it is often necessary to utilize a discrete form of the system. 
     Various discretization methods can be employed for this purpose, including the forward Euler method and Runge-Kutta methods.
   \section{Koopman Operator Theory}
   Consider the following discrete-time nonlinear system:
   \begin{equation}\label{Koopman}
   x_{k+1} = f(x_k),
   \end{equation}
   where $x_k$ is a state vector in the manifold $\mathcal{M} \subset \mathbb{R}^n$ on which $x_k$ evolves. Here, $k \in \mathbb{Z}$ denotes discrete time, and $f: \mathcal{M} \rightarrow \mathcal{M}$ is the evolution operator.
   
   To facilitate the analysis of this nonlinear system, we introduce the Koopman operator $\mathcal{K}$. According to \cite{koopman1931hamiltonian}, the evolution of the system described by (\ref{Koopman}) can be recast in an infinite-dimensional function space as follows:
   \begin{equation}\label{koppman1}
   (\mathcal{K} \zeta)(x_k) = (\zeta \circ f)(x_k) = \zeta(x_{k+1}),
   \end{equation}
   the Koopman operator $\mathcal{K}$ thus defines a new dynamical system $(\mathcal{H}, n, \mathcal{K})$. In this context, 
   $\zeta = [\zeta_1, \zeta_2, \cdots, \zeta_{\infty}] \in \mathcal{H}$ represents a set of lifting functions, where each 
   $\zeta_i: \mathbb{R}^n \rightarrow \mathbb{C}$ is termed an observable. Any scalar function of the state $x_k$ can qualify 
   as a lifting function. The symbol $\circ$ denotes the composition of the lifting function $\zeta$ with the evolution function $f$.
   
   The set of all such lifting functions forms an infinite-dimensional Hilbert space $\mathcal{H}$. The Koopman operator $\mathcal{K}: 
   \mathcal{H} \rightarrow \mathcal{H}$ is a linear operator that acts on these scalar-valued lifting functions $\zeta$ within $\mathcal{H}$.
    It is important to note that the Koopman operator maps functions of the state space to other functions of the state space 
    (i.e., $\mathcal{K}: \mathcal{H} \rightarrow \mathcal{H}$), rather than mapping states to states as $f: \mathcal{M} \rightarrow \mathcal{M}$ 
    \citep{budivsic2012applied}.
   
    Despite the fact that the Koopman operator $\mathcal{K}$ acts on lifting functions $\zeta$, making it potentially infinite-dimensional
    even when the original system $f$ is finite-dimensional, it is often possible to approximate $\mathcal{K}$ with a finite-dimensional 
    matrix while retaining a high degree of accuracy. This approximation is crucial for practical applications, as the exact finite-dimensional 
    representation of $\mathcal{K}$ is generally infeasible.
   
    According to \citet{williams2015data}, the infinite-dimensional representation (\ref{koppman1}) is approximated as follows in finite dimensions:
   \begin{equation}
   \zeta(x_{k+1}) = \mathcal{\tilde{K}} \zeta(x_k) + r_k,
   \end{equation}
   where $\zeta \in \mathbb{R}^N$,  $N \gg n$. The matrix $\mathcal{\tilde{K}} \in \mathbb{R}^{N \times N}$ represents a finite-dimensional approximation of the Koopman operator $\mathcal{K}$, and $r_k \in \mathcal{H}$ is the residual term, capturing the approximation error.
   
   This formulation enables the use of linear techniques to analyze and control the original nonlinear system, providing a powerful framework for understanding complex dynamical behaviors.
   

\subsection{Dynamical System with Input}
Consider the dynamical discrete nonlinear system with input:
\begin{equation}\label{system1}
x_{k+1} = f(x_k, u_k),
\end{equation}
where $x_k \in \mathcal{M} \subseteq \mathbb{R}^n$ represents the state vector, and $u_k \in \mathcal{N} \subseteq \mathbb{R}^p$ represents the input vector. 
The evolution of the state is governed by the nonlinear function $f: \mathcal{M} \times \mathcal{N} \rightarrow \mathcal{M}$. 
To analyze and control this nonlinear system, we introduce the Koopman operator $\mathcal{K}$. Specifically, the Koopman operator $\mathcal{K}$ acts on observable functions 
$\zeta$ defined on the state space. According to \cite{koopman1931hamiltonian}, the Koopman operator for the system in (\ref{system1}) is defined as:
\begin{equation}\label{koppman_u}
(\mathcal{K} \zeta)(x_k, u_k) = (\zeta \circ f)(x_k, u_k),
\end{equation}
where $\zeta(x_k, u_k) = [\zeta_1(x_k, u_k), \zeta_2(x_k, u_k), \cdots, \zeta_{N+p}(x_k, u_k)]$ is a vector of lifting functions, 
and each $\zeta_i: \mathbb{R}^n \times \mathbb{R}^p \rightarrow \mathbb{R}$ belongs to a space of observables $\mathcal{H}$.

The main objective of this work is to construct a finite-dimensional approximation of the Koopman operator, denoted as $\mathcal{\tilde{K}}$, and then use this approximation to predict the future states of the system. To achieve this, we define an extended state vector as follows:

\begin{align} \label{extendstate}
\xi_k = \begin{bmatrix} 
x_k \\ u_k 
\end{bmatrix},
\end{align}
the evolution of $\xi_k$ can be described by:
\begin{align} \label{extendevolves}
\xi_{k+1} = F(\xi_{k}) = \begin{bmatrix} 
f(x_k, u_k) \\ \mathcal{T} u_k 
\end{bmatrix},
\end{align}
where $\mathcal{T} u_k\in\mathbb{R}^p$. 
Consequently, equation (\ref{koppman_u}) can be rewritten as:
\begin{equation}\label{koppman2}
(\mathcal{K} \zeta)(\xi) = (\zeta \circ f)(\xi).
\end{equation}

To obtain the finite-dimensional approximation $\mathcal{\tilde{K}}$, we employ the EDMD algorithm \citep{williams2015data}. 
This algorithm uses a collection of data $(\xi_{k+1}, \xi_k)$ that satisfies
 $\xi_{k+1} = F(\xi_{k})$ to find the optimal approximation $\mathcal{\tilde{K}}$ by solving:
\begin{equation} \label{finiteK}
\min_{\mathcal{\tilde{K}}} \sum_{j=1}^{N} \| \zeta(\xi_{k+1}) - \mathcal{\tilde{K}} \zeta(\xi_{k}) \|_2^2.
\end{equation}
By applying the Koopman theory, we can express the nonlinear system (\ref{system1}) in the following linear form:
\begin{align} \label{liftmodel}
z_{k+1} &= Az_k + Bu_k, \nonumber \\ 
\hat{x}_k &= Cz_k,
\end{align}
where $z_k = \phi(x_k) = [\phi_{1}(x_k), \phi_{2}(x_k), \ldots, \phi_{N}(x_k)]^{T} \in \mathbb{R}^N$, and $\hat{x}_k$ is the prediction 
of $x_k$. The matrices $A \in \mathbb{R}^{N \times N}$, $B \in \mathbb{R}^{N \times p}$, and $C \in \mathbb{R}^{n \times N}$ need to be identified.
 We assume that the vector of lifting functions can be expressed as $\zeta(\xi_k) = \phi(x_k) + u_k$, leading to:
\begin{align} \label{linearpreditor}
\mathbf{\zeta}_{i}(\xi) = \begin{bmatrix} 
z_k \\  u_k 
\end{bmatrix}.
\end{align}
The equation (\ref{finiteK}) can be reformulated as:
\begin{equation}
\min_{A,B} \sum_{j=1}^{K} \| \phi(x_{k+1}) - A \phi(x_{k}) - B u_k \|_2^2.
\end{equation}
The optimal matrix $C$ can be computed as:
\begin{equation}
\min_C \sum_{j=1}^{K} \| x_j - C \phi(x_j) \|_2^2.
\end{equation}
For simplicity, let $\phi_{i} = x_i$ for $i = 1, 2, \ldots, n$, making $C = [I, 0]$, where $I$ is an $n \times n$ identity matrix.

This theoretical framework allows for the application of linear control techniques to nonlinear systems, providing a powerful tool for 
analyzing and controlling complex dynamical systems. The use of EDMD to approximate the Koopman operator enables the practical implementation 
of this approach in real-world scenarios.
\subsection{Finding the estimated Koopman operator}
Assuming the following set of data
\begin{align} \label{data}
X &= \begin{bmatrix} x_1 & x_2 &\dots & x_L \end{bmatrix},\nonumber \\
Y &= \begin{bmatrix} y_1 &  y_2 &\dots & y_L\end{bmatrix}, \nonumber\\
U &= \begin{bmatrix} u_1 & u_2& \dots & u_L \end{bmatrix},
\end{align}
where $y_k=f(x_k,u_k)$, this method does not impose the requirement for variables 
$x_i$ and $x_{i+1}$  to adhere to equal time intervals, nor does it necessitate that they originate from the same dynamic trajectory. 
This flexibility allows for the application of a broader dataset in analysis, free from the constraints of strict continuity typically 
associated with traditional time-series data.

After preparing the data, it is essential to apply the lifting function to map the data$(X,Y)$into the lifted space$(\bar{X},\bar{Y})$. There is no need to increase 
the dimensionality of $U$ since predicting $U$ is not required. 
The optimation problem is 
\begin{equation}\label{OptimationAB}
  \min_{A, B} \| \bar{Y} - A \bar{X} - BU \|_F,
\end{equation}
where $ \bar{Y}=[\phi(y_1),\phi(y_2),...,\phi(y_L)]$, $\bar{X}=[\phi(x_1),\phi(x_2),...,\phi(x_L)]$,$U=[u_1,u_2,...,u_L]$.

Tikhonov regularization, which adds a penalty based on the Frobenius norm of the matrix in question within a linear regression framework, 
can be employed to enhance the condition (\ref{OptimationAB}). This involves adjusting the EDMD cost function by incorporating
a regularization term.
\begin{eqnarray}\label{Tikhonov}
  J(U;\alpha) &=&\frac{1}{L}\|\bar{Y}-A\bar{X} - BU \|_F+\frac{\alpha}{L}\|[A,B]\|_{F}^{2}\nonumber\\
   &=&\frac{1}{L}trace((\bar{Y}-A\bar{X}-BU)(\bar{Y}-A\bar{X}-BU)^{T})\nonumber\\
   &&+\frac{\alpha}{L}trace([A,B][A,B]^{T})
\end{eqnarray}

\section{Koopman operator theory based MPC control}
Unlike nonlinear MPC, which involves solving challenging nonconvex optimization problems and relies heavily on local solutions, 
the data-driven approach reduces computational demands. In this section, we detail the design of MPC controllers for nonlinear systems using Koopman operator theory.
\subsection{MPC formulation}
To simultaneously consider the constraints on both the input (rotational speed) and the increment input 
(rate of change of rotational speed), the derived high-dimensional linear system$z_{k+1}=Az_k+Bu_k$ is transformed as follows:
\begin{align} \label{mpcd}
  \begin{bmatrix} 
    z_{k+1}\\u_k
  \end{bmatrix}=
  \begin{bmatrix} 
   A&B \\ 0&B 
  \end{bmatrix}
  \begin{bmatrix} 
    z_{k} \\ u_{k-1} 
    \end{bmatrix}+
    \begin{bmatrix} 
    0 \\ B 
    \end{bmatrix}\Delta u_{k-1},
  \end{align}
here, let $\bar{z}_{k}=[z_{k},u_{k-1}]^{T}$,$\bar{A}=[A,B ;0,B]$,$\bar{B}=[0,B]^{T}$. Then the MPC formulation is described as follows:
\begin{align}
	\min_{\Delta u_k} \quad J = &\sum_{k=0}^{Nh-1} (((\hat{x}_k-y_r)^T Q_u (\hat{x}_k-y_r)) + \Delta u_k^T R \Delta u_k)\nonumber\\
  &+((\hat{x}_{Nh}-y_r)^T Q_N (\hat{x}_{Nh}-y_r))\nonumber\\
		\text{s.t.}& \quad 
	\begin{cases}
    \bar{z}_{k+1} = \bar{A}\bar{z}_k + \bar{B}\Delta u_k,  \\ 
    \hat{x}_k = C\bar{z}_k,\quad \hat{x}_{\min} \le \hat{x}_k \le \hat{x}_{\max},\\
		u_{\min} \le u_k \le u_{\max},\quad\Delta u_{\min} \le \Delta u_{k}\le \Delta u_{\max}, 
	\end{cases}  \label{mpc2}
\end{align}
where $Q_u$, $R$, $Q_N$ are the  weight matrices, $Nh$ is the prediction horizon, $\bar{z}_k$ is the modified lifting vector, 
$\hat{x}_{\min}$ and $\hat{x}_{\max}$ are cosntraints on the output vector, $u_{\min}$ and $u_{\max}$ are the constraints of RPM,
 $\Delta u_{\min}$ and $\Delta u_{\max}$ are the constraints of increment input.

{\bf Remark 2:}
  Different from the work in \citet{korda_linear_2018}, this study considers 
  the constraints on both the input and increment input.
  According to \cite{korda_linear_2018}, the number of constraints remains unaffected by the dimensionality of the lifting function. 
  So, once we've figured out the nonlinear 
  mapping $z_0 = \phi(x_k)$, solving (\ref{mpc2}) doesn't cost much more than solving a standard linear MPC problem for the same prediction horizon. This remains 
  true even if both setups have the same number of control inputs and the state-space dimension is $n$ ranter than $N$.
\section{Simulations and results}
\subsection{Simulation in MATLAB}
The proposed approach is implemented to govern the dynamical system as outlined in equation (\ref{forwardspeed}). The system parameters are specified as follows:
 $m=146.471kg$, $X_{\dot{v}}=-4.876161$, $X_{|v|v}=-6.2282$, $t=0.1$, $\rho=1000kg/m^3$, $D=0.2m$, $\alpha_1=0.2$, $\alpha_2=0.1$, and $\omega=0.1$. 

An MPC controller is designed by utilizing exclusively input-output data, without knowing the model knowledge. 
The lifting-based predictor, as described by equation (\ref{liftmodel}), is derived by discretizing the scaled dynamics with the Runge-Kutta fourth-order 
method over a sampling period of $T_s = 0.01$ seconds. A simulation of 1000 trajectories is conducted, with each trajectory encompassing 100 sampling periods, 
equivalent to 1 second. The control input for each trajectory is a uniformly distributed random signal within the range of $[-50, 50]$.

The initial conditions for these trajectories are randomly selected from a uniform distribution over the interval $v \in [-0.5, 0.5]$. This data collection 
yields matrices $X$ and $Y$ of dimensions $ 1 \times 10^5$ and a matrix $U$ of size $1 \times 10^5$.Notably, $X$ and $Y$ are composed of the underwater vehicle's velocity $u$, while $U$ consists of the propeller's speed $s$.
The lifting functions $\phi$ are chosen to 
represent the state itself, that is, $\phi_1 = v$. The centers of the gauss functions
\footnote{The gauss function with center at $x_0$ is defined by $\phi(x)=e^{\frac{||x-x_0||^{2}}{2}}$.} are randomly selected from a uniform distribution 
over the interval $[-1, 1]$. Consequently, the dimension of the lifted state-space is set to $N = 5$.

\begin{figure}
    \centering
    \begin{subfigure}[b]{0.45\textwidth}
        \includegraphics[width=\textwidth]{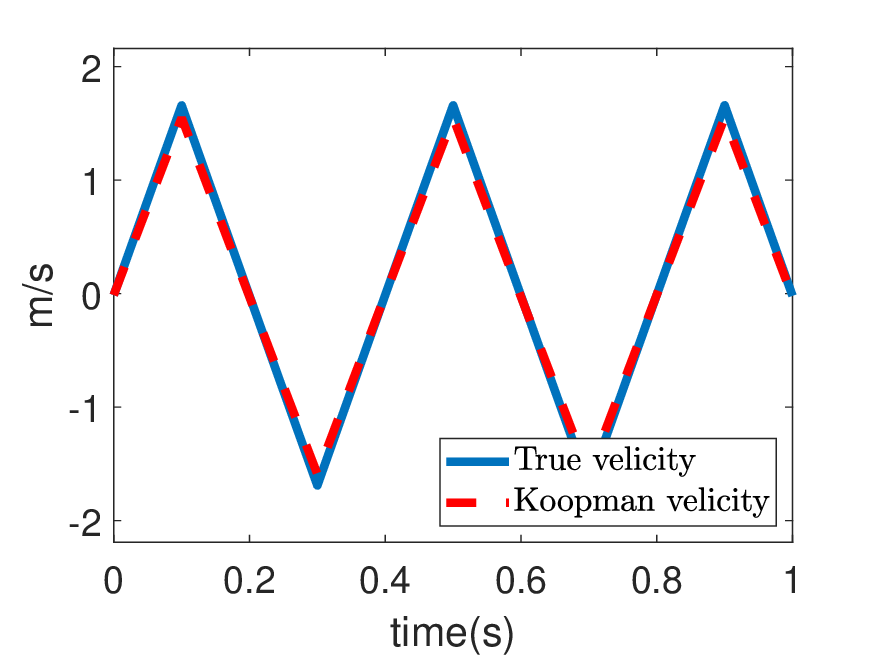}
        \caption{$v_0=0$}
        \label{fig:comparation}
    \end{subfigure}
    ~ 
    \begin{subfigure}[b]{0.45\textwidth}
        \includegraphics[width=\textwidth]{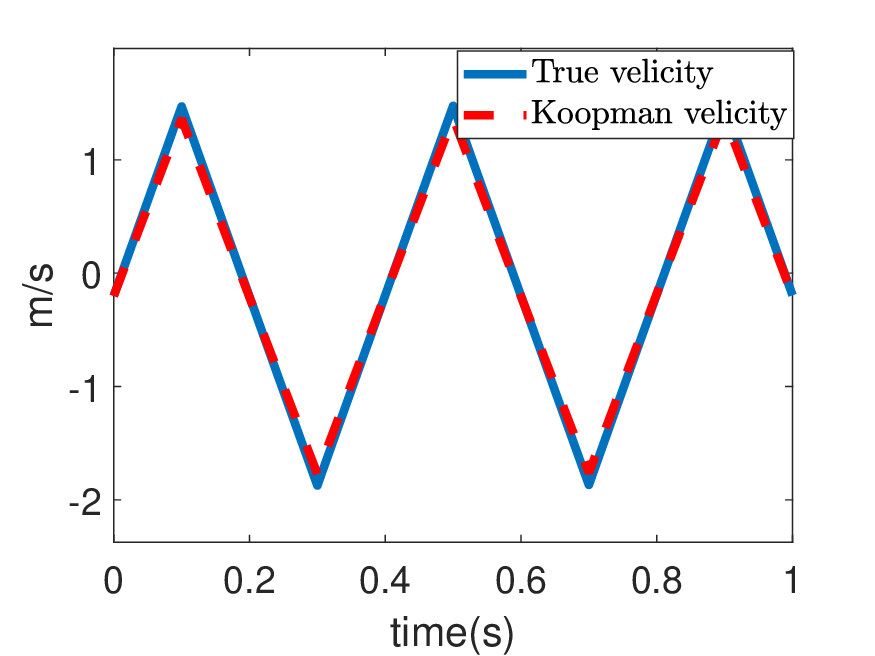}
        \caption{$v_0=-0.1$}
        \label{fig:comparation2}
    \end{subfigure}
    \caption{Prediction comparison for different initial value}\label{fig:comparestates}
\end{figure}

Fig.~\ref{fig:comparestates} illustrates the comparision between the actual system states and the states predicted by the Koopman operator, starting from two randomly selected 
initial conditions within the range $v \in [-0.5, 0.5]$ ($v_0^1=0$,$v_0^1=0.1$). The control signal $s(t)$ is depicted as a periodic wave with a period of 0.1 second 
and the magnitude is 40. Fig. 1 shows that the lifting-based Koopman predictor predicts the true velocity well.

By applying the Tikhonov algorithm (\ref{Tikhonov}) with collection data $X$, $Y$ and $U$, we can obtain the linear matrix expression
 (\ref{liftmodel}). 
Following this, MPC algorithm is utilized to control the underwater vehicle's speed by adjusting the propeller speeds.
 Here, set $C=[1,0,...,0]$, seclect the cost function matrices as 
 $Q_N=Q_u=2000$ and $R=0.01$. The prediction horizon is set to $Nh = 10$.  In the simulation, 
the input  $s$ is imposed the constraint as $[-50, 50]$,  the increment input is restricted as $[-20, 20]$, and a piece-wise reference $y_r(t)$ is tracked.
 
\begin{figure}[t]
  \centering
  \resizebox{11cm}{7cm}{\includegraphics{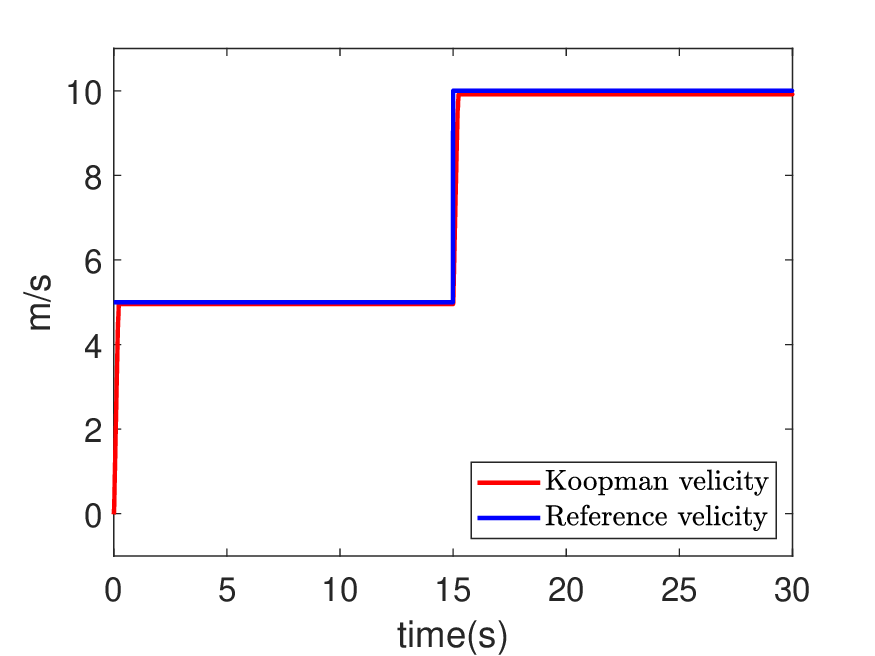}}
  \caption{MPC control for the refering tracking }\label{track}
  \end{figure}

\begin{figure}[t]
    \centering
    \resizebox{11cm}{7cm}{\includegraphics{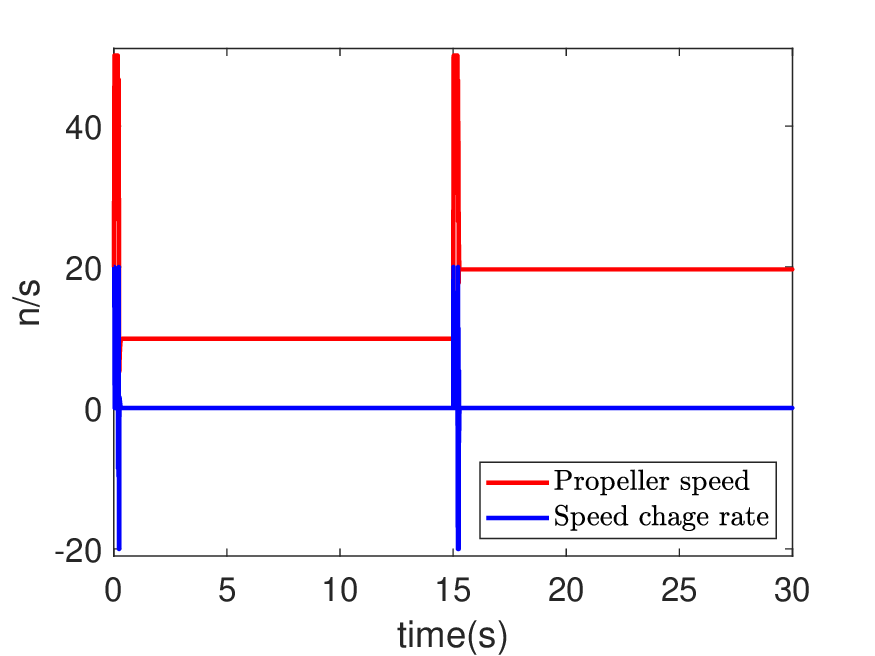}}
    \caption{The input and input change rate signal}\label{input}
    \end{figure}

In Fig.~\ref{track}, it is shown that even the maximum speed  and the increment speed of the propeller are constrained, 
the vehicle can track a given reference speed with the proposed Koopman MPC algorithm.
Fig.~\ref{input} illustrates that both the input  and increment input are constrained, 
in which the maximum speed of the propeller is 50 (PRM), and the increment speed is 20.
 
  {\bf Remark 3:} The system under study in this paper cannot be linearized locally due to its inherent complexity. 
  This scenario highlights the superiority of data-driven algorithms, which can effectively handle such complexities.
   The MPC algorithm is particularly advantageous as it can manage constraint issues. 
   By integrating Koopman operator theory with MPC, the control of unknown nonlinear systems can be effectively addressed, 
   while also managing constraints. This combination leverages the strengths of both theoretical frameworks to 
   achieve control solutions.
\subsection{Simulation in Gazebo}
\subsubsection{Simulation platform}
\begin{figure}
    \centering
    \begin{subfigure}[b]{0.45\textwidth}
        \includegraphics[width=\textwidth]{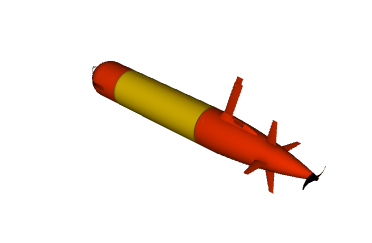}
        \caption{The Laruv underwater vehicle}
        \label{fig:Laruv}
    \end{subfigure}
    ~ 
    \begin{subfigure}[b]{0.45\textwidth}
        \includegraphics[width=\textwidth]{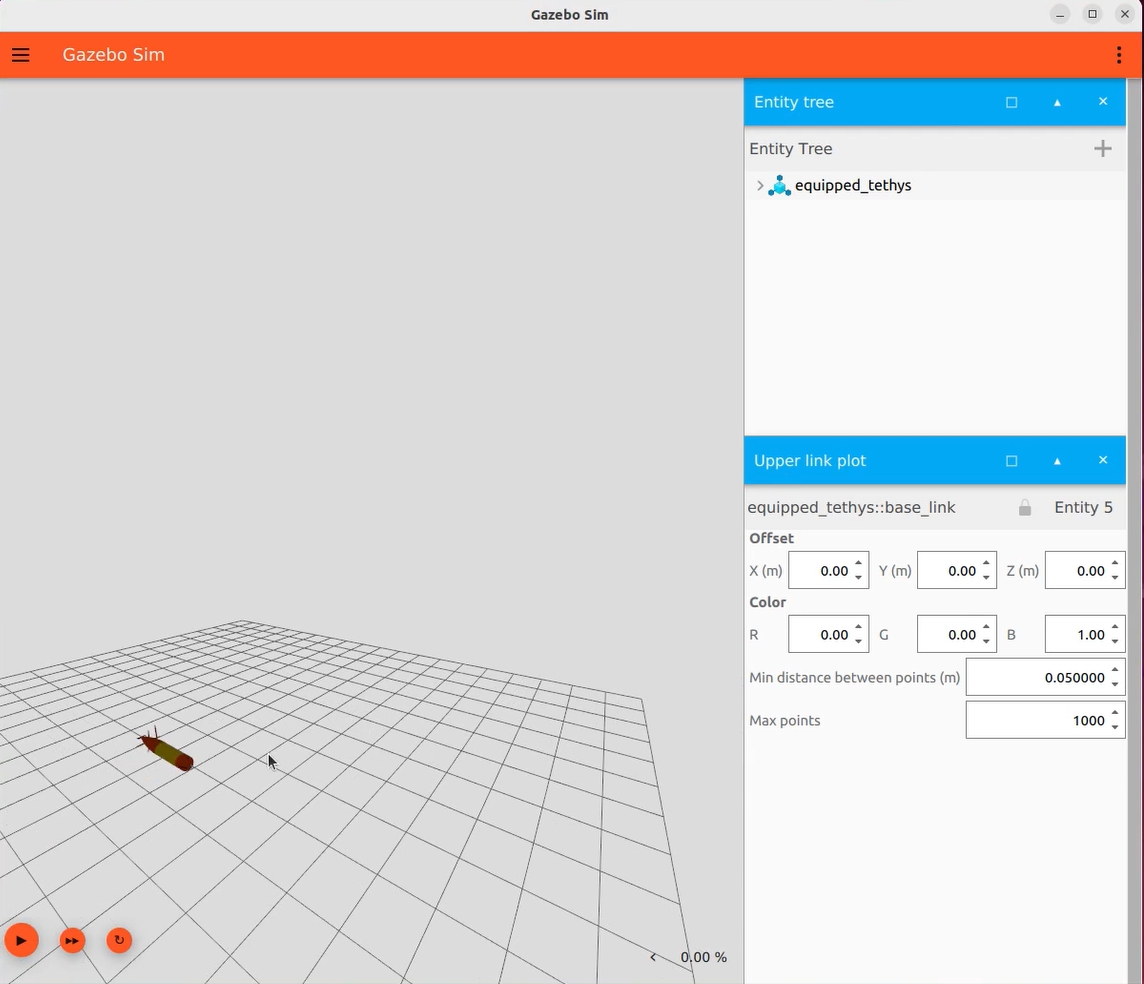}
        \caption{The simulated platform}
        \label{fig:platform}
    \end{subfigure}
    \caption{The simulated underwater vehicle and platform}\label{fig:simulatedplatform}
\end{figure}

In this study, we validate the proposed algorithm using open-source software, Gazebo and ROS. The operating system is 
Ubuntu 22.04(Intel(R) Core(TM) i7-7820HQ CPU @ 2.90GHz, 16G RAM), with Gazebo 
version Harmonic\footnote{https://gazebosim.org/docs/harmonic/getstarted/} and ROS version Iron\footnote{https://docs.ros.org/en/iron/Installation.html}. The communication between Gazebo and ROS is 
facilitated by the ROS-Gazebo bridge\footnote{https://gazebosim.org/docs/latest/ros2\_integration/}.
The underwater vehicle model used for simulation is based on the model proposed in \cite{player2023concept}, with all parameters configurable 
through the plugin files. 
LRAUV Sim1 is based on the new Gazebo, which has been entirely 
rewritten and is fundamentally different from the classic version of Gazebo. This updated simulator incorporates Fossen’s 
equations \citep{fossen2011handbook} to model hydrodynamics, similar to the approaches detailed in previous studies by \citet{zhang2022dave}.
This setup provides a more realistic test environment for AUV applications and algorithm design.
Fig. \ref{fig:simulatedplatform} illustrates the experimental simulator, where the vehicle's movement is powered by a stern thruster. There are four
rudders: two vertical rudders control the AUV's steering, while two horizontal rudders manage its ascent and descent. The speed of the AUV is 
controlled by adjusting the thruster's velocity. 

\subsubsection{Simulation Process}
Fistly, acquiring propeller speed and UAV's velicity data. Using the rosgz\_bridge package, the topics controlling the propeller
 and the 3D orientation of the underwater vehicle in Gazebo are linked with 
the publisher and subscriber topics in ROS. This integration allows us to obtain information about the vehicle, including propeller speed and 
vehicle orientation, within the ROS system. The propeller speed, randomly generated within the range of $[-10, 10]$, is published by a ROS 
publisher. Concurrently, a subscriber records the vehicle's velocity. By running the simulation for a certain period, we obtain data on 
propeller speed and vehicle velocity. In this experiment, the data collection frequency is set to $10 Hz$.
Secondly, we use the collected data and the proposed algorithm to derive a high-dimensional linear matrix. This step is implemented in MATLAB 2022b.
Finally, the computed matrices $\{A~B~C\}$ are loaded into ROS. Within a subscriber, the MPC algorithm is implemented using the scipy.optimize library.
 This algorithm calculates the propeller speed input based on the vehicle's velocity data from Gazebo. The calculated speed is then 
 sent to  Gazebo through a publisher which adjusts the  vehicle velocity. The vehicle is tasked with tracking a piecewise continuous reference 
 signal. 
 For this experiment, we set $Q_s=Q_u=2000$ and $R=0.01$. The input propeller speed $s$ is constrained within 
$[-150,150]$, with the increment input restricted to $[-50,50]$. A piecewise reference $y_r(t)$is tracked. The lift function and simulation 
settings are identical to the simulation in MATLAB.

\begin{figure}[t]
  \centering
  \resizebox{11cm}{6cm}{\includegraphics{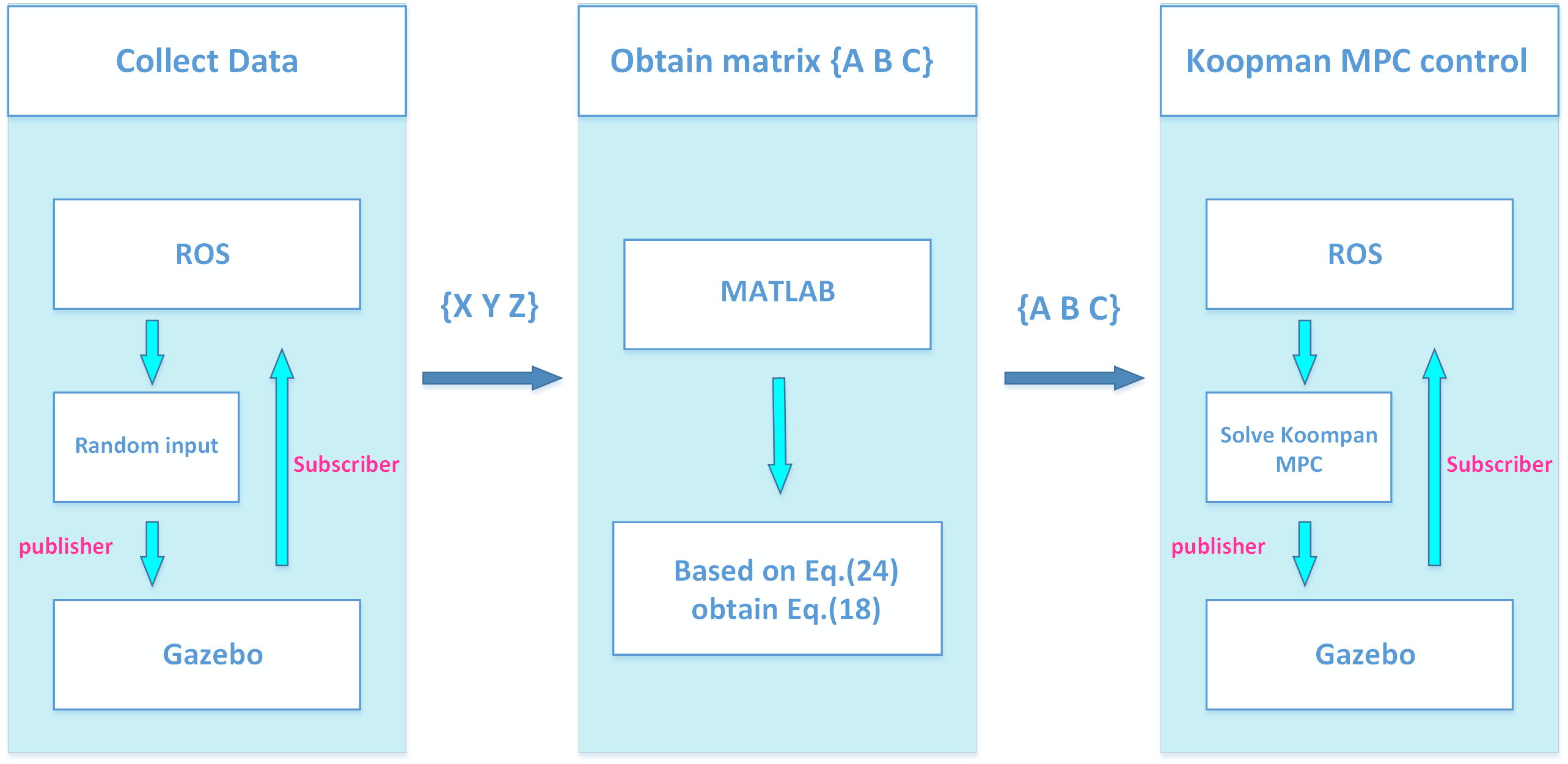}}
  \caption{The experiment process}\label{Gazeboprocess}
  \end{figure}
Fig.~\ref{Gazeboprocess} shows the process of the experiment.
\begin{figure}[t]
  \centering
  \resizebox{11cm}{7cm}{\includegraphics{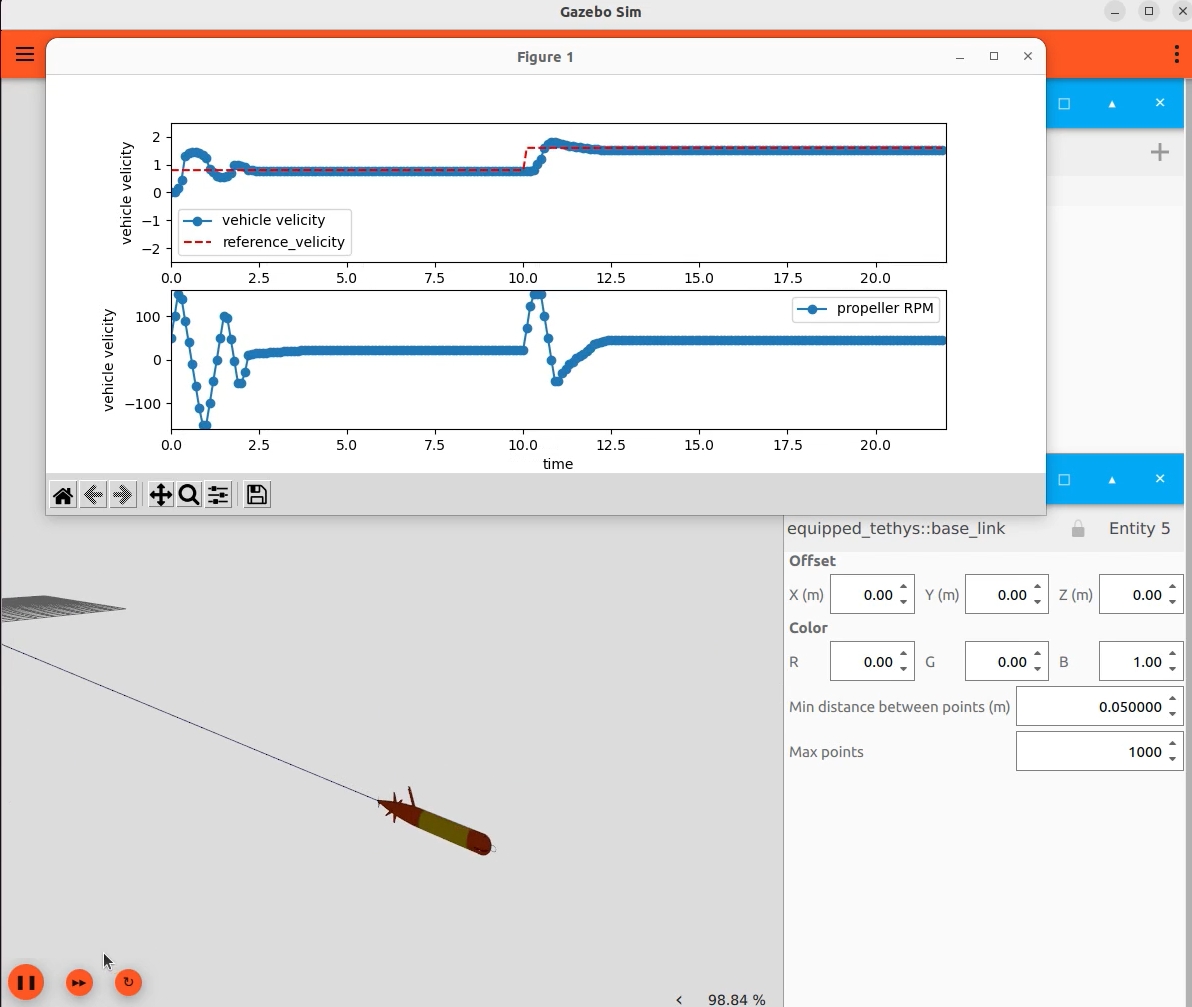}}
  \caption{The experiment result}\label{result}
  \end{figure}
  Fig.~\ref{result} demonstrates the vehicle's speed tracking using the proposed algorithm, with dynamic plots providing a more intuitive display of the 
  process. As shown in the figure, the vehicle effectively follows the given reference signal even when the target speed changes, 
  despite the constraints on propeller speed. This performance highlights the superiority of the MPC algorithm. 

\section{Conclusion}
This paper introduces a control strategy for AUVs that integrates a data-driven Koopman MPC. The approach adeptly manages the 
multivariable nonlinear dynamics of AUVs, addressing state, input, and increment input constraints to ensure operational stability
 and performance. Validated through a ROS2 and Gazebo-based platform, our method shows effectiveness in complex underwater environments, 
 offering a significant advancement for AUV control systems. The proposed strategies pave the way for more efficient and safer underwater
  exploration, with potential for further development and real-world application.

\section{CRediT authorship contribution statement}  
{\bf Zhiliang Liu:}  Original draft, Methodology, Project administration, Formal analysis, Data curation, Conceptualization. 
{\bf Xin Zhao:} Investigation, Review and editing. {\bf Peng Cai:} Visualization, Validation, Methodology. 
{\bf Bing Cong:} Validation, Review and editing.

\section {Declaration of competing interest}
The authors declare that they have no known competing financial interests or personal relationships that could have appeared to
 influence the work reported in this paper.

 \section{Data availability}
 Data will be made available on request. The experiment video can be found
  at \href{https://www.bilibili.com/video/BV1Ss421u7Qg/?spm_id_from=333.1007.top_right_bar_window_history.content.click&vd_source=f5a4c31759452d87ec6ee431fb397aaa}{here}.
 
 \section{Acknowledgment}
 This work was supported by Natural Science Foundation of China under grant 61873137. 


  \bibliographystyle{elsarticle-harv} 
  \bibliography{ThrusterMpc.bib}



\end{document}